\documentstyle[epsfig]{aipproc}

\begin{document}

\newbox\grsign \setbox\grsign=\hbox{$>$} \newdimen\grdimen \grdimen=\ht\grsign
\newbox\simlessbox \newbox\simgreatbox
\setbox\simgreatbox=\hbox{\raise.5ex\hbox{$>$}\llap
     {\lower.5ex\hbox{$\sim$}}}\ht1=\grdimen\dp1=0pt
\setbox\simlessbox=\hbox{\raise.5ex\hbox{$<$}\llap
     {\lower.5ex\hbox{$\sim$}}}\ht2=\grdimen\dp2=0pt
\def\simgreat{\mathrel{\copy\simgreatbox}}
\def\simless{\mathrel{\copy\simlessbox}}
\newbox\simppropto
\setbox\simppropto=\hbox{\raise.5ex\hbox{$\sim$}\llap
     {\lower.5ex\hbox{$\propto$}}}\ht2=\grdimen\dp2=0pt
\def\simpropto{\mathrel{\copy\simppropto}}

\def\ts{\thinspace}

\title{Evidence for a 304-day Orbital \\ Period for GX~1+4} 

\author{Jo\~ao Braga, Marildo G.\ Pereira and Francisco J.\ Jablonski}
\address{Divis\~ao de Astrof\'\i sica, Instituto Nacional de Pesquisas
Espaciais,\\ CP~515, 12201--970, S\~ao Jos\'e dos Campos, Brazil}

\maketitle

\begin{abstract}
  In this paper we report strong evidence for a $\sim$ 304-day
  periodicity in the spin history of the accretion-powered pulsar
  GX\ts 1+4 that is very likely to be a signature of the orbital
  period of the system. Using BATSE public-domain data, we show a
  highly-significant periodic modulation of the pulsar frequency from
  1991 to date which is in excellent agreement with the ephemeris
  proposed by Cutler, Dennis \& Dolan in 1986\cite{cdd86}, which were
  based on a few events of enhanced spin-up that occurred during the
  pulsar's spin-up era in the 1970s.  Our results indicate that the
  orbital period of GX\ts 1+4 is 303.8$\pm$1.1\ts days, making it by
  far the widest low-mass X-ray binary system known. A likely scenario
  for this system is an elliptical orbit in which the neutron star
  decreases its spin-down rate (or even exhibits a momentary spin-up
  behavior) at periastron passages due to the higher torque exerted by
  the accretion disk onto the magnetosphere of the neutron star.
\end{abstract}


\section*{Introduction} 

GX~1+4 is a unique accretion-powered pulsar in a low-mass x-ray binary
system (LMXB). In the 1970s the pulsar exhibited a spin-up behavior
with a rate of $\dot{P} \sim -2$\,s/year, the hightest among all
persistent X-ray pulsars, and was one of the brightest and hardest
X-ray sources in the sky.  After an extended low-intensity state in
the early 1980s, GX~1+4 re-emerged in a spin-down state \cite{mak88}
and has produced occasional short-term variations of $\dot{P}$ ever
since. The optical counterpart is a M5~III giant star, V2116\,Oph, in
a rare type of symbiotic system \cite{gla73,dav77,charo97}.  The
identification was made secure by a ROSAT accurate position
\cite{pre95} and by the discovery of optical pulsations consistent
with the spin period of the neutron star \cite{nos97,per97}. In 1991,
BATSE initiated a continuous and nearly uniform monitoring of GX~1+4,
confirming the spin-down trend with occasional dramatic spin-up/down
torque reversal events \cite{cha96,chaetal97}. GX~1+4 has a much
longer (factor of $\sim$ 100) spin period than the other four known
LMXB accretion-powered pulsars and its orbital period has been known
to be at least one order of magnitude longer than the periods of the
other systems \cite{charo97}. Attempts to find the orbital period by
Doppler shifts of the pulsar pulse timing\cite{cha96} or optical lines
\cite{dav77,dot81,soo95} have both been inconclusive so far. Using a
small number of X-ray measurements carried out during the spin-up
phase of GX~1+4 in the 1970s, Cutler, Dennis \& Dollan \cite{cdd86}
produced an ephemeris for predicting periodical enhancements in the
spin-up rate of the neutron star and claimed that this could be due to
an elliptical orbit with a 304-day period. Here we report the
discovery of a 304-day modulation in the BATSE frequency data and
discuss its implications to the models for this source.

\section*{Data Analysis and Results} 

The frequency and the pulsed flux data between Julian Day (JD)
2448376.5 and 2451138.5 (i.e., 1991 April 29 to 1998 October 20) used
in this work were obtained from Chakrabarty \cite{cha96} and from the
BATSE public domain data. The 20--50~keV pulsed signals are extracted
from DISCLA 1.024s channel 1 data. 15-day mean values for the fluxes
and pulse frequencies of GX~1+4 were calculated for the entire
dataset.

\begin{figure}[t!] 
\centerline{\epsfig{file=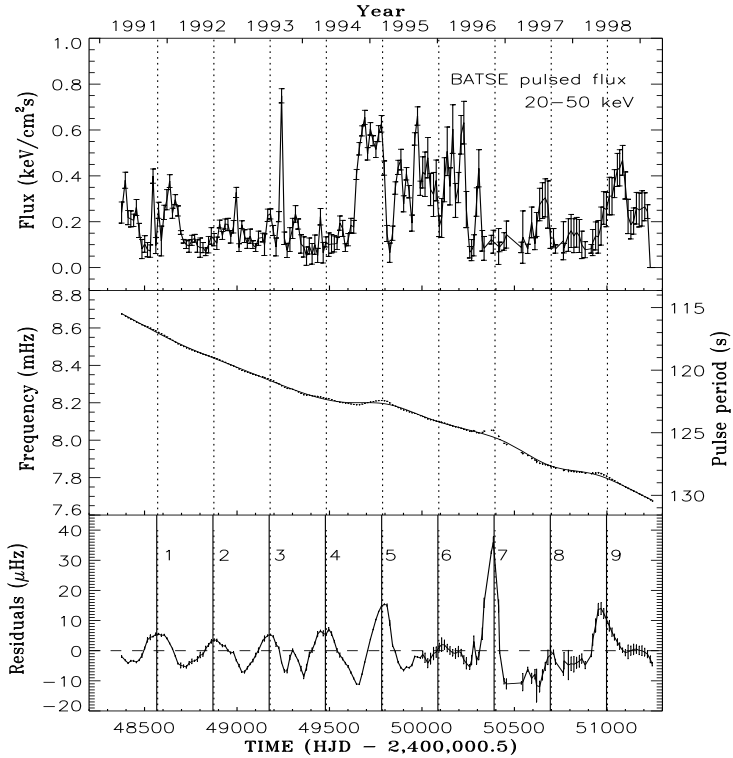,height=4.4in,width=5in}}
\caption{{\it Upper panel\/}: Light curve of the 20-50\ts keV
  pulsed flux of GX\ts 1+4 as measured by BATSE from 1991 to 1998;
  {\it middle panel\/}: GX\ts 1+4 frequency measurements by BATSE over
  the same period. The error bars are in general smaller than the size
  of the dots. The solid curve is a cubic spline fit to the data;
  {\it lower panel\/}: frequency residuals. The dotted vertical lines
  mark the times times predicted by the ephemeris calculated in this
  work, whereas the solid vertical lines show the predictions of
  Cutler, Dennis \& Dolan (1996). The events of positive residual
  frequency modulation are labeled for reference in the text.}
\label{fig1}
\end{figure}

A dataset of GX~1+4 residual pulsation frequencies was obtained from
the frequency history by subtraction of a standard cubic spline
function to remove low frequency variations in the spin-down trend.
The fitting points are mean frequency values calculated over suitably
chosen time intervals. The results of the spline fitting are fairly
insensitive to intervals greater than $\sim$ 200 days between fitting
points (we have used $\Delta t=215$~days). The pulsed X-ray flux,
frequency history and residual frequencies are shown in Figure
\ref{fig1} as functions of time.

\begin{figure}[t] 
\centerline{\epsfig{file=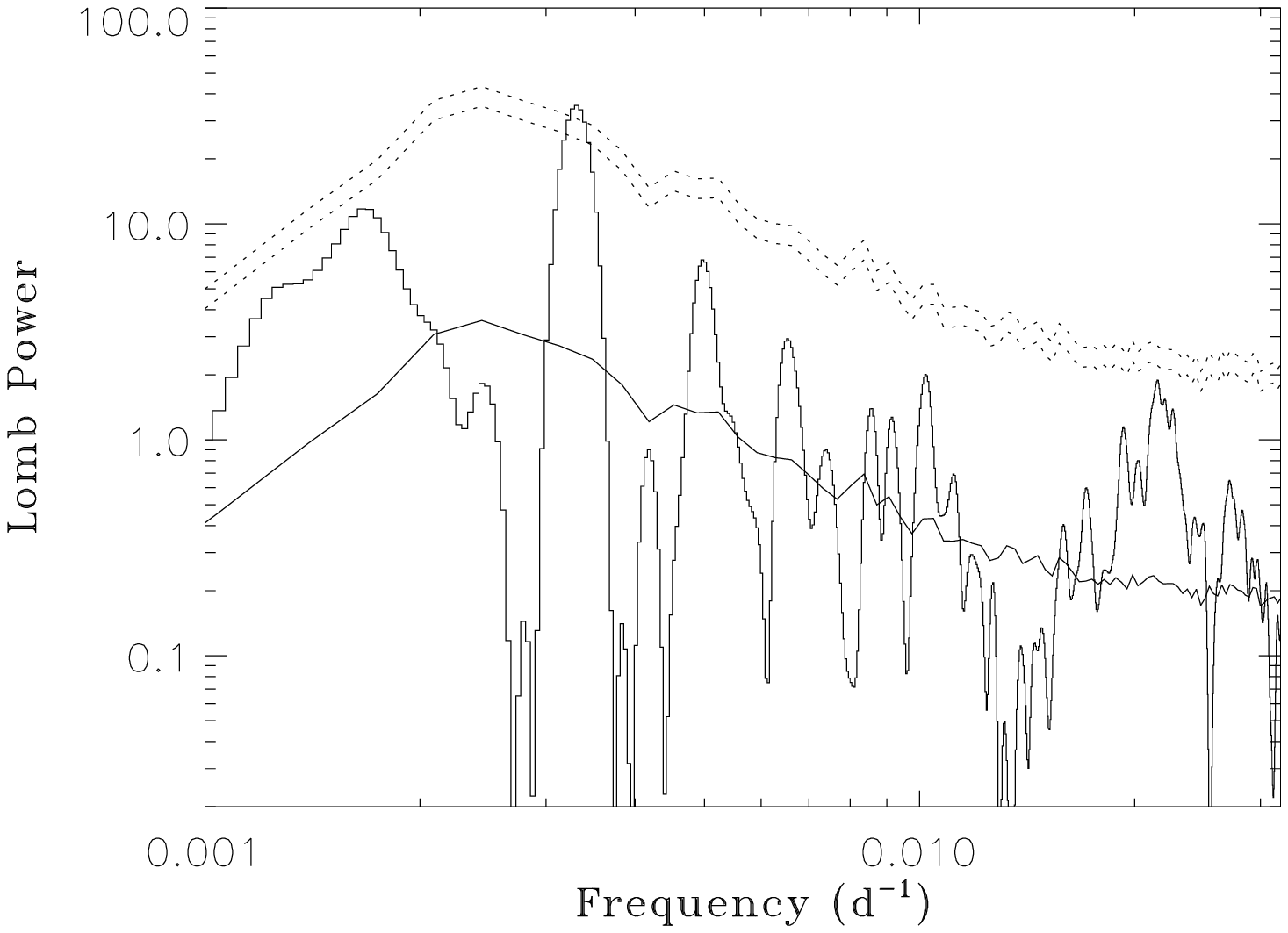,height=2.7in,width=4in}}
\caption{Lomb-Scargle periodogram of the frequency residuals of
GX\ts 1+4 from 1991 to 1998, represented by the histogram-type solid
line. The standard solid line is the mean of 1500 numerical
simulations carried out in order to calculate the significance level
of the detection. The upper dotted line indicates a significance level
of 0.001, whereas the lower dotted line indicates a significance level
of 0.01.}
\label{fig2}
\end{figure}

We have carried out a power spectrum analysis to search for
periodicities of less than 1000\,days in both the residual frequency
and the pulsed flux data. A Lomb-Scargle periodogram \cite{pre92},
suitable for time series with gaps, shows a significant periodic
signal at 302.0~days (Fig.\ \ref{fig2}) in the residual frequency time
series. The power spectrum shows a red noise with an approximate
power-law index index of $-$2. In order to estimate the statistical
significance of the detection, a series of numerical simulations of
the frequency time series with 1-sigma gaussian deviations were
performed \cite{per99}. The simulations show that the use of the 215-d
spline, besides providing an effective filter for frequencies below
$\sim 2\times 10^{-3} {\rm d}^{-1}$, does not produce power in any
specific frequency in the range of interest. By comparing the
amplitude of our 302-day peak with the local value obtained by the
mean of the numerical simulations, we obtain a statistical
significance of 99.98\% for the detection. Epoch folding the data
using the 302-day period yields a 1-$\sigma$ uncertainty of 1.7~days.

By analyzing the variation of the period of GX~1+4 during the spin-up
phase in the 1970s, Cutler, Dennis \& Dolan \cite{cdd86} proposed a
304\,-day orbital period and an ephemeris to predict the events of
enhanced spin-up: $T = {\rm JD\ 2,444,574.5} \pm 304\; n,$ where $n$
is an integer. This ephemeris is based on four events discussed by the
authors, whose existence was inferred from ad-hoc assumptions and
extrapolations of the observations.  The projected enhanced spin-up
events derived from that ephemeris for the epochs contained in the
BATSE dataset, represented as solid vertical lines in the lower panel
of Fig.\ \ref{fig1}, are in excellent agreement with the BATSE reduced
spin-down and spin-up events. The BATSE dataset is obviously
significantly more reliable than the one given by Cutler, Dennis \&
Dolan \cite{cdd86} since it is based on 9 well-covered events measured
with the same instrument as opposed to the 4 events discussed by those
authors. The striking agreement of their ephemeris with the BATSE
observations is very conspicuous and give a very strong support to the
claim that the orbital period of the system is indeed $\sim$ 304 days.
Taking integer cycle numbers, with the $T0$ epoch of Cutler, Dennis \&
Dolan \cite{cdd86} as cycle $-23$, and performing a linear
least-squares fit to the frequency residuals seen in the lower panel
of Fig.\ \ref{fig1}, we find that the following ephemeris can
represent the time of occurrence $T$ of the maxima in the frequency
residuals: $T = {\rm JD\ 2,448,571.3} (\pm 3.2) \pm 303.8 (\pm 1.1) \;
n$, where $n$ is any integer. The events predicted by the above
ephemeris are shown as vertical dotted lines in the three panels of
Fig.\ \ref{fig1}. The value of $303.8\pm 1.1$~days for the orbital
period is consistent with the one obtained through power spectrum
analysis performed on the BATSE data, which gives further support for
the period determination.

\section*{Discussion}

In the 1970s, when the measurements used by Cutler, Dennis \& Dolan
\cite{cdd86} were carried out, the source was in a spin-up extended
state. They proposed that the periodic occurrence of enhanced spin-up
events was due to the fact that the system was in a elliptical orbit
and the periastron passages would occur when $\dot{P}$ is maximum, as
expected in standard accretion from a spherically expanding stellar
wind. However, it is widely accepted today that the system has an
accretion disk.  Since the neutron star is currently spinning-down,
the radius at which the magnetosphere boundary would corotate with the
disk is probably smaller than the magnetosphere radius.  Since the
pulse period is $\sim$ 120~s and the luminosity is typically $\lesssim
10^{37}$~erg/s, it can be shown \cite{per99} that the period is
probably close to the equilibrium value, for which the two radii are
equal. This allows spin-down to occur even though accretion continues,
the centrifugal barrier not being sufficiently effective \cite{whi88}.
Assuming that the elliptical orbit is the correct interpretation for
the presence of the modulation, the mass accretion rate (and hence the
luminosity) should increase as the neutron star approaches periastron.
The spin-down torque then gets smaller and the neutron star
decelerates at a slower rate \cite{per99}. Occasionally, due to the
highly variable mass loss rate of the red giant, the neutron star will
{\it spin-up\/} for a brief period of time during periastron, as
observed in the BATSE frequency curve in events 5, 7 and 9. According
to this picture, one would expect an increase in X-ray luminosity at
periastron.  Although this is only marginally indicated in the BATSE
pulsed flux light curve, it should be pointed out that total flux data
from the ASM/RXTE for the epoch MJD 50088 to 51044 does not correlate
significantly with the BATSE pulsed flux, indicating that the pulsed
flux may not be a good tracer of the accretion luminosity in this
system. Furthermore, the periodic $\sim 5\mu$Hz excursions in the
residual frequency would lead to very low-significance variations in
the X-ray flux measured by the ASM \cite{per99}.

An alternative interpretation for the observed modulation would be the
presence of oscillation modes in the red giant star.  However, the
stability of the infrared magnitudes of V2116\ts Oph \cite{charo97}
preclude it from being a long-period variable, since these stars
undergo regular $\gtrsim$~1~mag variations in the infrared
\cite{whi87}. 

We conclude by pointing out that, given the 304-day orbital period and
the spectral and luminosity characteristics of V2116\ts Oph, it can be
shown that the companion in this system is probably not filling its
Roche lobe and the accretion disk forms from the slow, dense stellar
wind of the red giant \cite{per99}. A more thorough covering of the
X-ray luminosity of the system, with high sensitivity and spanning
several cycles, will be very important to test the elliptical model
for GX\ts 1+4.


We thank Dr.\ Bob Wilson from NASA Marshall Space Flight Center for
gently providing us BATSE frequency and flux data on GX\ts 1+4.  M.\
P.\ is supported by a FAPESP Postdoctoral fellowship at INPE under
grant 98/16529-9. J.\ B.\ thanks CNPq for support under grant
300689/92-6. F.\ J.\ acknowledges support by PRONEX/FINEP under grant
41.96.0908.00.

\end{document}